\documentclass[journal]{IEEEtran}

\date{}
\usepackage{cite}
\usepackage{url}
\usepackage{amsmath}
\usepackage{array}
\usepackage{graphicx}
\usepackage[dvipsnames]{xcolor}
\usepackage{caption} % For subfigures
\usepackage{subcaption} % For subfigures
\usepackage{times}
\usepackage[none]{hyphenat}
\usepackage{url}
\usepackage{latexsym}
\usepackage{indentfirst}
\usepackage{graphicx}
\usepackage{booktabs}

\author{Noah Rhodes and Line Roald\\
Dept. of Electrical and Computer Engineering\\ 
 University of Wisconsin-Madison \\
 Madison, WI\\ 
 {\underline{nrhodes@wisc.edu}, \underline{ roald@wisc.edu}} \\
 }
 
\begin{document}
% \chardef\_=`_
\newcommand{\com}[1]{\begin{code}{#1}\end{code}\nonumber}
\newcommand{\upperlim}[1]{\overline{#1}}
\newcommand{\lowerlim}[1]{\underline{#1}}

\newcommand{\bus}{\mathcal{B}}
\newcommand{\busid}{i}
\newcommand{\busidtwo}{j}

\newcommand{\gen}{\mathcal{G}}
\newcommand{\genid}{g}
\newcommand{\busgen}{\gen^{\bus_\busid}}

\newcommand{\branch}{\mathcal{L}}
\newcommand{\branchid}{l}
\newcommand{\busbranch}{{\branch}^{\bus_\busid}}

\newcommand{\storage}{\mathcal{E}}
\newcommand{\storageid}{e}
\newcommand{\busstorage}{{\storage}^{\bus_\busid}}

\newcommand{\demand}{\mathcal{D}}
\newcommand{\demandid}{d}
\newcommand{\busload}{{\demand}^{\bus_\busid}}

\newcommand{\shunt}{\mathcal{S}}
\newcommand{\shuntid}{s}
\newcommand{\busshunt}{{\shunt}^{\bus_\busid}}

\newcommand{\power}{P}
\newcommand{\reactive}{Q}
\newcommand{\timstep}{t}

\newcommand{\pgen}{\power_{\genid,\timstep}}
\newcommand{\pgenupper}{\upperlim{\boldsymbol{\power}_{\genid,\timstep}}}
\newcommand{\pgenlower}{\lowerlim{\boldsymbol{\power}_{\genid,\timstep}}}
\newcommand{\qgen}{\reactive_{\genid,\timstep}}
\newcommand{\qgenupper}{\upperlim{\boldsymbol{\reactive}_{\genid,\timstep}}}
\newcommand{\qgenlower}{\lowerlim{\boldsymbol{\reactive}_{\genid,\timstep}}}

\newcommand{\pstorage}{\power_{\storageid,\timstep}}
\newcommand{\pstorageupper}{\upperlim{\boldsymbol{\power}_{\storageid,\timstep}}}
\newcommand{\pstoragelower}{\lowerlim{\boldsymbol{\power}_{\storageid,\timstep}}}
\newcommand{\qstorage}{\reactive_{\storageid,\timstep}}
\newcommand{\qstorageupper}{\upperlim{\boldsymbol{\power}_{\storageid,\timstep}}}
\newcommand{\qstoragelower}{\lowerlim{\boldsymbol{\power}_{\storageid,\timstep}}}

\newcommand{\pbranch}{\power_{\branchid,\timstep}}
\newcommand{\qbranch}{\reactive_{\branchid,\timstep}}

\newcommand{\pload}{\boldsymbol{\power}_{\loadid}}
\newcommand{\qload}{\boldsymbol{\reactive}_{\loadid}}

\newcommand{\damaged}{\times}

\newcommand \nr[1] {{\color{black}#1}}
\newcommand \lr[1] {{\color{black}#1}}

\title{The Role of Distributed Energy Resources in \\ Distribution System Restoration}

\maketitle

\begin{abstract}
With increasing levels of distributed energy resources (DERs) connected to the grid, it is important to understand the role that DERs can play in post-disaster restoration. In this paper, we propose a \lr{two-step} optimization method to identify and implement an optimal restoration schedule under different DER operating scenarios. We investigate how the presence \lr{and geographical distribution} of DERs change the optimal restoration order, and assess the impacts on customers with and without DERs. In our case study using the IEEE 123 single phase \nr{distribution} system, we find that optimal restoration order changes significantly when DERs are concentrated in one part of the grid. We also observe that the presence of DERs generally reduces the energy not served across all customers and can help prioritize grid reconnection of customers without DERs. 
\end{abstract}

\section{Introduction}

Recovering the power grid quickly after a disaster is imperative to maintain quality of life and limit economic impacts. Extended outages can have severe impacts, as seen in Puerto Rico after hurricanes Irma and Maria, where customers on average went without electricity for 84 days \cite{NEJMsa1803972}, or in California during the Public Safety Power Shutoffs in 2019 where millions of customers, including medical baseline customers \cite{sotolongo_bolon_baker_2020}, were without power for several days \cite{DeEnergization}. 
Many methods to improve disaster resilience in electric power systems have been proposed, including preparedness and response to earthquakes \cite{xu2007optimizing}, wildfires \cite{rhodes2020balancing}, and hurricanes \cite{kavousi2018stochastic}. 
In the context of post-disaster recovery, several studies have considered combined problems such as grid sectionalization and repair planning \cite{qiu2017integrated}, two stage problems considering component delivery and repair scheduling \cite{arif2017power}, and joint optimization of repair crew routing and repair scheduling \cite{van2011vehicle}.
Others consider pre-disaster hardening of distribution grid components to limit outage size \cite{barnes2019resilient, wang2015research}, 
or leverage microgrid technology to provide black-start power \cite{castillo2013microgrid} or alleviate impacts on islanded distribution systems \cite{li2014distribution}.

\lr{In the future, Distributed Energy Resources (DERs), such as rooftop solar PV and electric vehicles, will likely play an increasingly important role in providing power during emergencies. 
DERs provide an opportuntity }
to support household load while the electrical system is damaged, and alleviate some of the impact of a power outage to their owners \cite{belding2020will}.  
If the DERs can power not just a local building, but also create a microgrid within an isolated region, this may improve the network capabilities even further \cite{gupta2019achieving}.
However, the presence of DERs also raises new questions about equity, fairness and energy access among different customers in the restoration process. The owners of DERs like rooftop solar PV or an electric vehicle have a higher income than the national average \cite{lukanov2019distributed, farkas2018environmental}, while other customers may either lack the means or ability to install DERs (e.g., tenants in rented apartments). 
This may lead to an uneven distribution of DERs across the grid, 
as neighborhoods of different socio-economic background experience different penetration of DER technology. \nr{As DER penetrations increase, it is increasingly important to analyze the effects of such differences on customers both during normal operations and emergency situations.}

\nr{In this context,} this paper aims to explore the impact of DERs in aiding power system restoration, and analyze how the presence of DERs affects DER owners and non-owners within the system. \lr{We examine how these results vary with different scenarios for DER operation, including scenarios where DERs either (i) disconnect (or stop working for other reasons) when the grid experiences an outage,} (ii) power an individual home (a home microgrid) or (iii) exchange power with neighboring properties (a community microgrid). We also consider how the solutions differ in scenarios where DERs are distributed uniformly throughout the distribution feeder versus scenarios where DERs clustered within a specific 
region.  
In particular, we are interested in understanding how the presence of DERs interacts with 
the question of \emph{which parts of the grid to restore first}. 
We study this question in the context of the restoration ordering problem (ROP) \cite{van2011vehicle}, which is a multi-time step optimization problem that identifies an optimal restoration plan to minimize customer impacts. 
The ROP is very challenging to solve as it requires simultaneous consideration of a large number of damaged components and their respective on/off status at each time step, 
leading to a large-scale, mixed-integer non-linear problem.
An important modeling aspect is the choice of the power flow formulation, which can have significant impact on both the complexity and accuracy of the problem \cite{rhodes2020powermodelsrestorationjl}.
To obtain numerically tractable problem formulations, researchers either do not consider power flow \cite{xu2007optimizing}, use simplified power flow representations \cite{arif2017power,van2011vehicle}, or employ the full AC power flow representation \cite{qiu2017integrated,barnes2019resilient}).
\lr{Here, we propose to use a simplified DC power flow model in the mixed-integer ROP problem to identify a restoration order. Once the restoration order is known, we solve a continuous, multi-period AC optimal power flow that determines the active and reactive power dispatch at each time step. While not guaranteed to be optimal, this approach has been shown to provide a good trade-off between problem scalability (system size and number of damaged elements) and solution quality \cite{rhodes2020powermodelsrestorationjl}. }

In summary, the contribution of this paper is as follows.
First, we formulate a model for restoration planning in distribution grids under different DER operating scenarios. To enable computational tractability, the restoration is modeled as a two-step process, where the first step identifies an optimal \nr{restoration plan} and the second step simulates the implementation. We consider three different modes 
where the DERs are either unavailable, or operate as either a home or community microgrid. Second, we analyze the impact of different DER operational scenarios on the optimal restoration plan, and assess outcomes for customers with and without DERs.

The remainder of the paper is organized as follows. 
Section \ref{sec:problem} discusses the modeling and assumptions of our problem formulation, and the mathematical formulation of the optimization problems. Section \ref{casestudy} introduces the test case, then Section \ref{results} presents the results and analysis of this work.  Finally, Section \ref{conclusion} concludes the paper.

\section{Grid Restoration With DERs}\label{sec:problem}
In the following, 
we present our model set-up and performance criteria, 
and the mathematical formulation for the two stages of restoration.

We consider a \nr{radial} distribution network where the sets of all lines, nodes, and load demand are denoted as $\branch, \bus,$ and $\demand$. 
\lr{The sources of power include the substation, as well as utility-owned and customer-owned DERs, which are represented by the set $\gen$.
In the development of our restoration plan, we consider damage to any utility owned asset, including lines, buses and utility-owned DERs.} The subset of damaged elements are marked with superscript $\damaged$, i.e. $\branch^\damaged, \gen^\damaged, \bus^\damaged$, where
\nr{
$\gen^\damaged$ only contain utility-owned DERs and the substation.}
We denote demand at nodes with and without DERs as $\demand^{DER}$ and $\demand^{0}$, and the substation, which also serves as a reference node, is denoted as $\bus^{ref}$. 
The set of time periods is given by $\mathcal{T}$.
Individual generators, nodes, demands and time steps are indexed by $g\in\gen, i\in\bus, d\in\demand$ and $t\in\mathcal{T}$, respectively. Lines are indexed by $(l,i,j)\in\branch$ where $l$ represents the line index, and $i$ and $j$ are the from and to nodes of each line. 
The subset of lines, generators and demands that are connected to a specific bus $i$, are denoted by $\branch^{\bus_i}, \gen^{\bus_i}, \bus^{\bus_i}, \demand^{\bus_i}$.
The decision variables such as the generator output $P_{\genid}$ are denoted in normal script.  
Constants such as line susceptance $\boldsymbol{b}$ are in bold, and upper and lower bounds are shown with an over and underline respectively, e.g., $\underline{\boldsymbol{\theta}}, \overline{\boldsymbol{\theta}}$.

\subsection{Performance Criteria and Objective}
In an optimization problem, one of the main modeling decision is the choice of the performance criterion that determines the objective function. 
Typical objective functions for restoration problems include minimizing average customer time without power \cite{xu2007optimizing}, minimize time until the last load is connected \cite{qiu2017integrated}, maximize power delivery during restoration \cite{kavousi2018stochastic,van2011vehicle,li2014distribution}, or
co-optimize maximum load-delivery and minimum restoration time \cite{arif2017power}. 
Different objective functions that optimize for different measures of performance may lead to different optimal restoration sequences and  
will prioritize of certain customers above others. These priorities may be explicit (e.g., prioritization of critical loads such as hospitals \cite{nejad2019distributed}) or implicit (e.g., faster connection of neighborhoods with hospitals). 

In this paper, we will consider two different performance criteria for our evaluation of the restoration plan. The first criterion is to maximize the amount of energy served in to customers while the grid is being repaired, which is used to define our objective function. We express this objective as 
\begin{equation}
	 \max \quad \sum_{t \in \mathcal{T}}  \sum_{\demandid \in {\demand}} x_{\demandid,t}   \boldsymbol{P}^D_{\demandid} \boldsymbol{\Delta}_t~. \label{eq:rop_objective}
\end{equation}
Here, $\boldsymbol{\Delta}_t$ is the duration of each time step. The parameters $\boldsymbol{P}^D_d$ represent the power demand at each node $d\in\mathcal{D}$. The continuous decision variable $0\leq x_{\demandid,t}\leq 1$ represents the the fraction of load provided to demand $d$.
This objective function prioritizes repairs that restore large loads quickly, 
which implicitly prioritizes customers who consume more power. 

Our second performance criterion is the average time to reconnect a customer load to the substation.
We evaluate the average reconnection time for two different groups of customers, namely customers \emph{with} DERs (which may have either full or partial access to power due to their local DER systems) and customers \emph{without} DERs. The respective average reconnection times $T^{DER},~T^{0}$ are defined as 
\begin{equation}
    T^{DER} = \frac{\sum_{d\in\mathcal{D}^{DER}}T_d}{|\mathcal{D}^{DER}|}, \quad
    T^{0} = \frac{\sum_{d\in\mathcal{D}^{0}}T_d}{|\mathcal{D}^{0}|},
    \label{eq:reconnection}
\end{equation}
where $T_d$ is the time of connection for the demand $d$.

\subsection{DER Operating Scenarios}\label{sec:setup_der}
We consider three different scenarios for DERs during a grid restoration scenario: 

\noindent\emph{0) Base Case} 
In our base case scenario, the DERs are omitted from the restoration planning and/or assumed to not provide power during the implementation of the restoration plan. This scenario represents a situation in which the DERs are either unavailable due to damage from the disaster or not able to provide reliable power (e.g., \nr{DERs that lack the capability to continue running during a blackout}, PV panels during night time hours or a battery that has been drained of power due to a long outage duration). In this scenario, the DERs are omitted from the model.

\noindent\emph{1) Home Microgrid} 
In our home microgrid scenario, the DERs are assumed to power the individual homes where they are connected, but do not exchange any power with the grid and do not share excess power with nearby nodes. In this scenario, we model the DERs as a reduction in the load at each node. Specifically, we define the new load as
\begin{equation}
	 \boldsymbol{D}_{\demandid} = \max (0.01\boldsymbol{D}_{\demandid}^{org}, \boldsymbol{D}_{\demandid}^{org}-P^{G}_d). \quad \forall \demandid \in \demand \label{eq:load_reduction}
\end{equation}
Here, $\boldsymbol{D}_{\demandid}^{org}$ is the original load (without DER power) and $P^G_d$ represents the power provided by the DER at node $d$.
The lower threshold for demand on nodes with DERs is 1\% of the original load. 
This threshold ensures that the load has a non-zero priority of reconnection.

\noindent\emph{2) Community Microgrid} 
In our community microgrid scenario, the DERs power not only the node where they are connected, but also exchange power with the grid and the neighboring nodes (i.e., they operate as a microgrid). In this scenario, we model the DERs as dispatchable generation which can adjust their power output within certain limits,
\begin{subequations}
\label{eq:rd_gen_limits}
\begin{align}
    &\boldsymbol{z_{\genid,t}}\underline{\boldsymbol{P}_{\genid}} \le P_{\genid,t}^G \le \boldsymbol{z_{\genid,t}} \overline{\boldsymbol{P}_{\genid}} && \forall {\genid} \in \mathcal{G}, \forall t \in \mathcal{T} \label{eq:der_p_gen_limits}\\
    &\boldsymbol{z_{\genid,t}}\underline{\boldsymbol{Q}_{\genid}} \le Q_{\genid,t}^G \le \boldsymbol{z_{\genid,t}} \overline{\boldsymbol{Q}_{\genid}} && \forall {\genid} \in \mathcal{G}, \forall t \in \mathcal{T}. \label{eq:der_q_gen_limits}
    \end{align}
\end{subequations}
Here, $P_{\genid,t}^G, Q_{\genid,t}^G$ represents active and reactive power generation at each time step $t$, with respective upper and lower bounds $\underline{\boldsymbol{P}^{G}_{\genid}}, \ \overline{\boldsymbol{P}^{G}_{\genid}}$ and   $\underline{\boldsymbol{Q}^{G}_{\genid}} \ \overline{\boldsymbol{Q}^{G}_{\genid}}$. The variable $z_{\genid,t}\in\{0,1\}$ represents the repair status of the DER; if the DER is operating, $z_{\genid,t}=1$ and the DER is subject to the above generation limits; if the DER is damaged and not yet repaired, $z_{\genid,t}=0$ and the power output is constrained to be $0$.

\subsection{The Two Stages of Restoration: Planning and Implementation}
Given the above performance criteria and DER operation scenarios, we study two stages of the post-disaster restoration of a distribution grid:

\noindent\emph{1) Restoration Ordering Problem (ROP)}
The first stage is to solve the Restoration Ordering Problem (ROP) to plan the sequence of repairs. 
This problem is very challenging to solve, 
as it requires the modeling of the binary repair decision $z$ as well as the power flow across multiple time-steps. To achieve tractability, we omit the voltage magnitude and reactive power from the problem and use a linear DC approximation of the power flow. \nr{ROP with DC Power Flow has been shown to be adequate for radial networks because the connectivity is more important than the voltage magnitude and reactive power constraints \cite{rhodes2020powermodelsrestorationjl}.}

\noindent\emph{2) Restoration Implementation Problem (RIP)}
The second stage simulates the implementation of the restoration plan, which we refer to as the Restoration Implementation Problem (RIP). In this problem, the restoration sequence is fixed (i.e., there are no more binary decision variables), and the problem becomes a continuous multi-period power flow problem. 
In this problem, we utilize the more detailed non-linear AC power flow formulation, which considers reactive power and voltage magnitudes. 

One important difference between the ROP and the RIP stages is that the RIP problem deploys a more accurate power flow formulation that considers voltage magnitudes and reactive power.
Another important difference is that the DER operating scenario assumed in the ROP may be inaccurate or outdated by the time the restoration plan is implemented. 
For example, when solving the ROP, the utility may not know whether the DERs themselves are also damaged, or if their availability is otherwise changed (e.g., a solar PV that produces less energy due to cloudy skies
). However, these changes could be accounted for when later solving the RIP.
By comparing the results of the ROP and the RIP, we can thus assess whether the optimal restoration plan is sensitive to such inaccuracies in the assumptions.

\subsection{Restoration Ordering Problem} \label{ROP}
We first present the \nr{ROP problem to determine} the order in which to restore damaged grid components.

\subsubsection{Component Repair Constraints}

The ROP is formulated as a multi-period optimization problem where the time periods $t\in\mathcal{T}$ are linked by the repair decisions $z_{i,t}\in\{0,1\}$.
Due to limitations on, e.g., available repair crews and equipment, there are typically limits on how many elements can be repaired in each time period. 
However, we can distinguish between the \emph{physical repair} of a component (e.g., replacing a downed conductor) and the \emph{re-energization} of the same component (i.e., when the component is taken back into operation). 
Here, we assume that although only a limited number of components can be repaired in each time step, it is possible to delay the re-energization of repaired components (e.g. if required to avoid infeasibility of the problem). If $z_{i,t}=1$, component $i$ is both repaired and energized. If $z_{i,t}=0$, the component is may be repaired, but is not yet energized.
The constraint on the number of re-energized components is thus expressed as an upper bound on the cumulative number of re-energized elements at each time step,
\begin{subequations}
\label{eq:repairs}
\begin{align}
    & \sum_{\genid \in \gen}z_{\genid,t} +
    \sum_{\branchid \in \branch}z_{{\branchid},t} +
    \sum_{\busid \in \bus}z_{\busid,t} \le \boldsymbol{R}_t && \forall t \in \mathcal{T}. \label{eq:repair_limit}
\end{align}
Here, we define $\boldsymbol{R}_t$ to allow 1 new repair in each time period, such that $\boldsymbol{R}_t = \boldsymbol{R}_{t-1}+1$. We also set $\boldsymbol{R}_{0}=0$, to represent that all damaged elements are initially unrepaired. \nr{Note that this upper bound could easily be adjusted to allow for more than one repair per time period. }
Further, we require that re-energized elements remain in operation after they have been repaired, i.e.,
\begin{equation}
    z_{k,t}\! \le \!z_{k, t+1} ~~~ \forall k \in \left\{ \gen^{\damaged}\!, \bus^{\damaged}\!, \branch^{\damaged}\!, \demand^{\damaged} \right\}, \forall t \in \mathcal{T}. \label{eq:repair_active}
\end{equation}
We also have to prevent components attached to a node from being re-energized before the node is back in operation. For example, a distribution line that is attached to a damaged node should not be re-energized before the node is also repaired and re-energized. This restriction is expressed as
\begin{align}
    &  z_{k,t} \leq z_{\busid,t} \quad \forall k \in \left\{\busgen, \busbranch, \busload \right\},
     \forall t \in \mathcal{T} \label{eq:repair_bus_comp} 
\end{align}
All devices must be repaired in the final time period,
\begin{equation}
    z_{k,t_{final}} = 1 \quad\quad \forall k \in \left\{ \gen^{\damaged}, \bus^{\damaged}, \branch^{\damaged}, \demand^{\damaged} \right\}. \label{eq:repair_all}
\end{equation}
\end{subequations}

\subsubsection{DER and Substation Constraints}

Power is provided from the substation and/or from local DER resources. The substation is modeled as a generator, 
\begin{equation}
     \underline{\boldsymbol{P}^G_{\genid}} \le P_{\genid,t}^G \le  \overline{\boldsymbol{P}^G_{\genid}} \quad \forall t \in \mathcal{T}~, \label{eq:gen_limits}
     \vspace{-0.5em}
\end{equation}
where the upper and lower bounds $\underline{\boldsymbol{P}^G_{\genid}}, \overline{\boldsymbol{P}^G_{\genid}}$ provide sufficient capacity to serve all the loads or absorb excess renewable energy as necessary.
The DER constraints are determined by the DER scenario, as described in Section \ref{sec:setup_der}. In the base case, the DERs are omitted. In the home microgrid scenarios, the DERs are modeled as a reduction in the load. In the community microgrid scenario, the DERs are treated as generation with active power limits determined by \eqref{eq:der_p_gen_limits}.

\subsubsection{DC Power Flow Constraints}

The DC Power Flow approximation neglects impact of reactive power and voltage magnitude variations, and assumes small angle differences. 
For the ROP problem, the DC power flow equations are given by
\begin{subequations}
\label{eq:powerflow}
\begin{align}
& \sum_{g\in\mathcal{G}^{\mathcal{B}_i}}\!\!\!P_{g,t}^G +\!\!\!\! \sum_{l\in\mathcal{L}^{\mathcal{B}_i}}\!\!\!P_{(\branchid,i,j),t}^L -\!\!\!\! \sum_{d\in\mathcal{D}^{\mathcal{B}_i}}\!\!\!\!x_{d,t} \boldsymbol{P}^D_d \!=\! 0 && \!\!\!\forall i \!\in\! \mathcal{B}  \label{eq:power_balance}
\\
& P_{(\branchid,i,j),t}^{L} \!\le\! \!-\boldsymbol{b}_{\branchid} (\theta_{i,t} \!-\! \theta_{j,t} \!+\! \overline{\boldsymbol{\theta}}(1\!-\!z_{\branchid,t})\!) && \!\!\!\!\!\!\!\!\! \!\!\!\!\!\!\!\forall (l,i,j) \!\in\! \mathcal{L} \!\! \label{eq:flow_limit1} \\
& P_{(\branchid,i,j),t}^{L} \!\ge\! \!-\boldsymbol{b}_{\branchid} (\theta_{i,t} \!-\! \theta_{j,t} \!+\! \underline{\boldsymbol{\theta}} (1\!-\!z_{\branchid,t})\!) && \!\!\!\!\!\!\!\!\! \!\!\!\!\!\!\!\forall (l,i,j) \!\in \!\mathcal{L} \!\label{eq:flow_limit2}\\
&-\boldsymbol{T_{\branchid}}z_{\branchid,t} \le  P_{{(\branchid},i,j),t}^{L} \le \boldsymbol{T_{\branchid}}z_{\branchid,t} && \!\!\!\!\!\!\!\!\!\!\!\!\!\!\!\!\!\forall (l,i,j) \!\in\! \mathcal{L} \!\label{eq:thermal_limit}\\
& \theta_i = 0 && \!\!\!\!\!\!\!\!\!\!\!\!\!\!\!\!\!\forall i \!\in\! \mathcal{B}^{ref} \!\! \label{eq:ref_angle} 
\end{align}
\end{subequations}
Here, \eqref{eq:power_balance} represents the nodal power balance. 
The constraints \eqref{eq:flow_limit1} and \eqref{eq:flow_limit2} express the power flow on each line $(l,i,j)\in\mathcal{L}$ as a function of the line suseptance $\boldsymbol{b}_{\branchid}$, the voltage angles $\theta_i$ and $\theta_j$ and the re-energization status of the line $z_{l,t}$. 
When $z_{\branchid,t}=1$,  \eqref{eq:flow_limit1} and \eqref{eq:flow_limit2} form an equality that represent the standard DC power flow. When $z_{\branchid,t}=0$, the angles on each side of the line must be decoupled from the power flow $P_{(l,i,j),t}^L$. 
This is achieved by choosing sufficiently large values of
$\overline{\boldsymbol{\theta}}$ and $\underline{\boldsymbol{\theta}}$,
\nr{
for example by following the method suggested in}
\cite{hijazi2017convex}. 
\lr{We note that the formulation \eqref{eq:flow_limit1}, \eqref{eq:flow_limit2} assumes $\boldsymbol{b}_{\branchid}\geq 0$. 
If $\boldsymbol{b}_{\branchid} < 0$, we flip the sign of the inequality in \eqref{eq:flow_limit1}, \eqref{eq:flow_limit2}.}
\nr{There exist other alternative models that compute a big-M value for the power flow rather than the angle differences, e.g.  \cite{kocuk2016DCOTS}.} Finally, \eqref{eq:thermal_limit} then enforces the power flow $P_{(l,i,j),t}^L$ to be $0$ when $z_{\branchid,t}=0$ and within the thermal limits when $z_{\branchid,t}=1$. Eq. \eqref{eq:ref_angle} sets the angle at the reference bus to zero. 

The above equations hold for a scenario where all lines in the system are damaged. However, we can adapt the equations by replacing the variable $z_{l,t}$ by a parameter $\boldsymbol{z_{l,t}}=1$ for un-damaged lines $l\in\mathcal{L}\setminus\branch^{\damaged}$.

\begin{table*}[]
    \small
	\renewcommand{\arraystretch}{1.75}
	\begin{center}
		\begin{tabular}{l}
			\toprule
			\vbox{ \begin{equation}
                \small \displaystyle 
    			P_{\branchid,\busid,\busidtwo,t}^L 
    			= \boldsymbol{z_{\branchid,t}}\left( \frac{\boldsymbol{g}+\boldsymbol{g_{fr}}}{\boldsymbol{t_m}^2} V_{\busid,t}^2 +  \frac{-\boldsymbol{g} \boldsymbol{t_r}+\boldsymbol{b} \boldsymbol{t_i}}{\boldsymbol{t_m}^2}\left( V_{\busid,t} V_{\busidtwo,t} cos\left(\theta_{\busid,t}-\theta_{\busidtwo,t}\right)\right) + \frac{-\boldsymbol{b} \boldsymbol{t_r}-\boldsymbol{g} \boldsymbol{t_i}}{\boldsymbol{t_m}^2}\left(V_{\busid,t}  V_{\busidtwo,t} sin\left(\theta_{\busid,t}-\theta_{\busidtwo,t} \right) \right) \right) \label{eq:rd_p_power_flow_fr}
			\end{equation}}\\ 
			\vbox{ \begin{equation}
    			\small \displaystyle P_{{\branchid},\busidtwo,\busid,t}^L = \boldsymbol{z_{{\branchid},t}}\left( \frac{\boldsymbol{g}+\boldsymbol{g_{to}}}{\boldsymbol{t_m}^2} V_{\busidtwo,t}^2 + \frac{-\boldsymbol{g} \boldsymbol{t_r}-\boldsymbol{b} \boldsymbol{t_i}}{\boldsymbol{t_m}^2}\left( V_{\busidtwo,t} V_{\busid,t} cos\left(\theta_{\busidtwo,t}-\theta_{\busid,t}\right)\right) + \frac{-\boldsymbol{b} \boldsymbol{t_r}+\boldsymbol{g} \boldsymbol{t_i}}{\boldsymbol{t_m}^2}\left(V_{\busidtwo,t}  V_{\busid,t} sin\left(\theta_{\busidtwo,t}-\theta_{\busid,t} \right) \right) \right)  \label{eq:rd_p_power_flow_to}
			\end{equation}}\\
			\vbox{ \begin{equation}
    			\small \displaystyle  Q_{{\branchid},\busid,\busidtwo,t}^L = \boldsymbol{z_{{\branchid},t}}\left(-\frac{\boldsymbol{b}+\boldsymbol{b_{fr}}}{\boldsymbol{t_m}^2}  V_{\busid,t}^2 - \frac{-\boldsymbol{b} \boldsymbol{t_r}-\boldsymbol{g} \boldsymbol{t_i}}{\boldsymbol{t_m}^2} \left( V_{\busid,t} V_{\busidtwo,t} cos\left(\theta_{\busid,t}-\theta_{\busidtwo,t}\right)\right) + \frac{-\boldsymbol{g} \boldsymbol{t_r}+\boldsymbol{b} \boldsymbol{t_i}}{\boldsymbol{t_m}^2} \left( V_{\busid,t} V_{\busidtwo,t} sin\left(\theta_{\busid,t}-\theta_{\busidtwo,t}\right)\right) \right) \label{eq:rd_q_power_flow_fr}
			\end{equation}}\\
			\vbox{ \begin{equation}
    			\small \displaystyle   Q_{{\branchid},\busidtwo,\busid,t}^L = \boldsymbol{z_{{\branchid},t}}\left(-\frac{\boldsymbol{b}+\boldsymbol{b_{to}}}{\boldsymbol{t_m}^2}  V_{\busidtwo,t}^2 -  \frac{-\boldsymbol{b} \boldsymbol{t_r}+\boldsymbol{g} \boldsymbol{t_i}}{\boldsymbol{t_m}^2} \left( V_{\busidtwo,t} V_{\busid,t} cos\left(\theta_{\busidtwo,t}-\theta_{\busid,t}\right)\right) + \frac{-\boldsymbol{g} \boldsymbol{t_r}-\boldsymbol{b} \boldsymbol{t_i}}{\boldsymbol{t_m}^2} \left( V_{\busidtwo,t} V_{\busid,t} sin\left(\theta_{\busidtwo,t}-\theta_{\busid,t}\right)\right) \right) \label{eq:rd_q_power_flow_to}
			\end{equation}}\\
			\bottomrule
		\end{tabular}
		\vspace{-5pt}
	\caption{AC power flow for all lines $(l,i,j) \in \mathcal{\branch}$, $t \in \mathcal{T}$}
	\vspace{-15pt}
	\label{tab:RIP_line_power}
	\end{center}
\end{table*}

\subsubsection{ROP Formulation}

The full \nr{ROP formulation} is given as follows:
\begin{subequations} \label{eq:rop}
\begin{align}
    &\max\limits_{x, z, P^G, P^L, \theta} && 
    \mbox{Energy Served to Customers
    \eqref{eq:rop_objective}} \nonumber\\[-8pt]
&\quad\quad  s.t. && \eqref{eq:repairs}, \eqref{eq:gen_limits}, \eqref{eq:der_p_gen_limits},\eqref{eq:powerflow} \nonumber 
\end{align}
\end{subequations}
We note that \nr{the ROP} is a large-scale, mixed-integer linear program that scales with the system size and the number of damaged components. For each damaged device, we need to consider an additional repair period, which requires a new set of constraints and variables to represent the power flow and operational status of that time period.

\subsection{Restoration Implementation Problem} \label{RIP}
The Restoration Implementation Problem (RIP) simulates the implementation of the restoration plan. 
In this problem, we obtain the optimal restoration order $z*$ from the ROP, and use those binary variables $z$ to define the set of status parameters $\boldsymbol{z}$. 
Further, we use an AC power flow model to model the effects of reactive power and voltage magnitudes \nr{and verify that an AC feasible power flow exists for this restoration plan.}

\subsubsection{Managing Voltage Magnitude Constraints} \label{ss:V_equations}

We introduce new variables $V_{\busid,t}$ to represent the voltage magnitudes for all buses $i\in\bus$ and time steps $t\in\mathcal{T}$. To keep voltage magnitudes within their upper and lower bounds $\overline{\boldsymbol{V}_{\busid}}, \underline{\boldsymbol{V}_{\busid}}$, we enforce the following constraints,
\begin{subequations}
\label{eq:rd_voltage_limits}
\begin{align}
    & V_{\busid,t} \leq \boldsymbol{z_{\busid,t}} \overline{\boldsymbol{V}_{\busid}} && \forall \busid \in {\bus}, t\in\mathcal{T} \label{eq:rd_v_mag} \\
    & V_{\busid,t} \geq \boldsymbol{z_{\busid,t}} \left( \underline{\boldsymbol{V}_{\busid}} - \tilde{V}_{{\busid},t} \right) && \forall \busid \in {\bus}, t\in\mathcal{T}, \label{eq:rd_v_vio}\\
    &  \tilde{V}_{{\busid},t} \geq 0 && \forall \busid \in {\bus}, t\in\mathcal{T}
\end{align}
\end{subequations}
Here, we observe that \eqref{eq:rd_v_mag} and \eqref{eq:rd_v_vio} together forces the voltage magnitude $V_{\busid,t}=0$ if $\boldsymbol{z_{\busid,t}}=0$. If  $\boldsymbol{z_{\busid,t}}=1$, the upper voltage bound is enforced by \eqref{eq:rd_v_mag} whereas \eqref{eq:rd_v_vio} enforces a soft lower bound on the voltage magnitude, with the variable $\tilde{V}_{\busid,t}$ representing the amount of violation. This soft constraint ensures feasibility if, e.g., an undamaged bus is isolated from the grid and lacks a power source to keep the voltage magnitude within bounds. To discourage violations of the lower voltage bound, we include a penalty in the objective function, i.e.,

\begin{equation}
	 \max \!\!\!\!\!\quad \sum_{t \in \mathcal{T}} \! \left( \sum_{\demandid \in {\demand}} x_{\demandid,t}  \boldsymbol{D}_{\demandid} - \sum_{\busid \in {\bus}} \tilde{V}_{{\busid},t}\right) \boldsymbol{\Delta}_t~. \label{eq:rd_objective}\\
\end{equation}

\subsubsection{AC Power Flow Constraints}

The AC power flow constraints for the active and reactive power flow $P^L_{(l,i,j),t}, Q^L_{(l,i,j),t}$ in both direction on a transmission line are expressed by the equations \eqref{eq:rd_p_power_flow_fr}-\eqref{eq:rd_q_power_flow_to} in Table \ref{tab:RIP_line_power}. These are the standard AC power flow equations, but modified to include the parameters $\boldsymbol{z}_{l,t}$ which represent the repair status of the lines. When $\boldsymbol{z}_{l,t}=0$, the power flow in both directions is zero. The $\boldsymbol{g}$ represents the line conductance magnitude, while $\boldsymbol{t_m}$ is the transformer ratio where the real and imaginary components are $\boldsymbol{t_r}$ and $\boldsymbol{t_i}$, respectively. 
The nodal power balance constraints for active and reactive power are given by
\begin{align}
    & \sum_{g\in\mathcal{G}^{\mathcal{B}_i}}P_{g,t}^G + \sum_{l\in\mathcal{L}^{\mathcal{B}_i}}P_{\branchid,i,j,t}^L - \sum_{d\in\mathcal{D}^{\bus_i}}x_{d,t} \boldsymbol{P}^D_d = 0    \label{eq:rd_p_power_balance}\\
    & \sum_{g\in\mathcal{G}^{\mathcal{B}_i}}Q_{g,t}^G + \sum_{l\in\mathcal{L}^{\mathcal{B}_i}}Q_{\branchid,i,j,t}^L - \sum_{d\in\mathcal{D}^{\bus_i}}x_{d,t} \boldsymbol{Q}^D_d = 0  \label{eq:rd_q_power_balance}
\end{align}
for all $i \in \mathcal{B}, t \in \mathcal{T} \nonumber$, where $\boldsymbol{Q}^D_d$ is the reactive power demand. 
We also include angle difference constraints
\begin{align}
    &  \boldsymbol{z_{{\branchid,t}}} \left(\theta_{\busid,t} \!-\! \theta_{\busidtwo,t}\right) \le \boldsymbol{z_{{\branchid,t}}} \overline{\boldsymbol{\theta}}_{{\busid},{\busidtwo}} && \!\! \forall ({\branchid,i,j}) \in \branch,  t \in \mathcal{T}, \label{eq:rd_flow_limit1}\\
    &  \boldsymbol{z_{{\branchid,t}}} \left(\theta_{\busid,t} \!-\! \theta_{\busidtwo,t}\right) \ge \boldsymbol{z_{{\branchid,t}}} \underline{\boldsymbol{\theta}}_{{\busid},{\busidtwo}} && \!\! \forall ({\branchid,i,j}) \in \branch,  t \in \mathcal{T}, \label{eq:rd_flow_limit2}
\end{align}
as well as thermal limits $\forall ({\branchid,i,j}) \in \branch,  t \in \mathcal{T}$
\begin{align}
    &  \sqrt{{P_{(l,i,j),t}^L}^2 + {Q_{(l,i,j),t}^L}^2}  \le \overline{\boldsymbol{T}}_\branchid \boldsymbol{z_{{\branchid,t}}} && \!\! . &   
    \label{eq:rd_thermal_limit}
\end{align}

\subsubsection{RIP Formulation}

Together, the objective and constraints form the following optimization problem,
\begin{subequations} \label{eq:rd}
\begin{align}
    &\max\limits_{\substack{x, P^G, Q^G, P^L  \\ Q^L, \theta, V, \tilde{V}}} &&  \mbox{Energy Served - Voltage Penalty \eqref{eq:rd_objective}} \nonumber\\
&\quad\quad s.t. && \eqref{eq:rd_gen_limits}, \eqref{eq:rd_voltage_limits},\eqref{eq:rd_p_power_flow_fr}-\eqref{eq:rd_thermal_limit} \nonumber
\end{align}
\end{subequations}
The problem is a multi-period AC optimal power flow with continuous decision variables.

\section{Case Study Set-Up} \label{casestudy}

We next describe the test case and implementation which is used to generate the results in the next section.

\subsection{Test Case}
We base our case study on the single phase version of the IEEE 123 bus distribution feeder \cite{bolognani2016} with 56 nodes, 54 lines and 52 loads. The total power demand is 3.49 MW, with an average demand 68 kW per load. The topology of the base system is shown in Fig \ref{fig:damage_scenario}, with the substation in dark blue, and buses in green. The size of the buses represent the amount of load demand.

\subsubsection{Damage scenario}

\begin{figure}[t]
    \centering
    \includegraphics[width=0.4\textwidth]{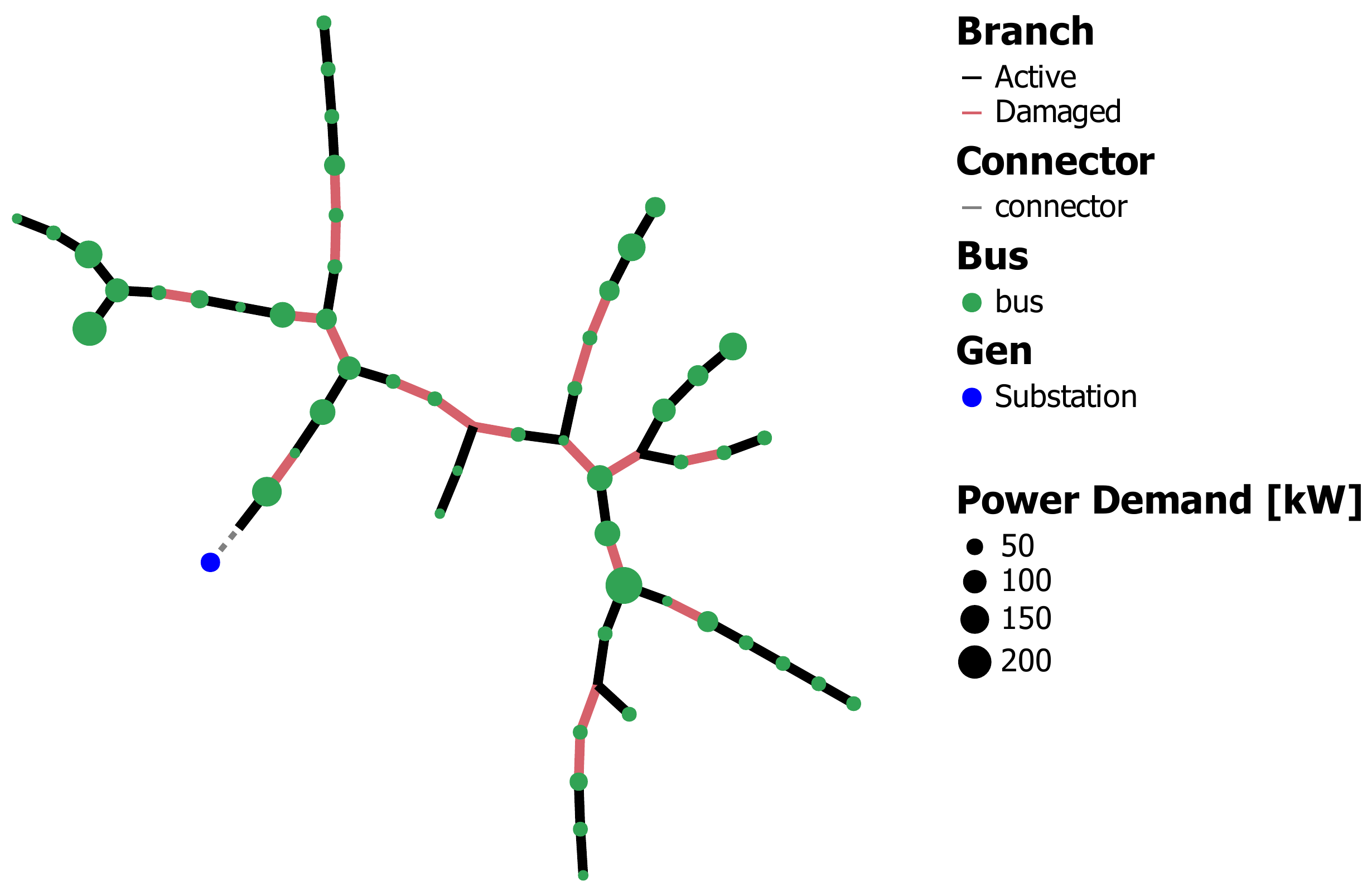}
     \caption{\small Damage scenario for the IEEE 123 single phase equivalent network. The undamaged branches are shown in black, the damaged branches are indicated in orange and the substation is indicated by the dark blue circle. The size of the nodes (in green) are scaled to indicate the power demand. The DERs are not shown in this figure.}
     \vspace{-10pt}
     \label{fig:damage_scenario}
\end{figure}

We consider a damage scenario arising from a severe wind storm, which causes damage to overhead lines in the distribution grid. We assume that 18 lines are damaged, as  indicated in orange in Fig \ref{fig:damage_scenario} and listed in Table \ref{tab:parameters}. The total number of time steps is $|\mathcal{T}|=19$, since we also include an initial time step to represent the original network before the repairs are started. We assume that each time step is an hour, and the total energy demand of the network during the restoration is thus $66.31$ MWh.

\subsubsection{Allocation of DERs}

To the base system, we add DERs with an active power rating $0\leq{P}^G_g\leq75$ kW and reactive power limits of $-50\leq{Q}^G_g\leq50$ kVAR. 
We compare two geographical distributions of DERs:\\[+2pt]
    \underline{\emph{Uniform DERs:}} 28 DERs are assigned uniformly throughout the grid. Their locations are shown in light blue in Fig \ref{fig:rop_result} b) and are listed in Table \ref{tab:parameters}. \\[+2pt]
    \underline{\emph{Clustered DERs:}} 16 DERs are added in a cluster on one region of the distribution system. Their locations are shown in light blue in Fig \ref{fig:rop_result} e) and are listed in Table \ref{tab:parameters}.\\[+2pt]
Note that both the number and locations of DERs are different in the two cases, implying a different DER penetration of the overall feeder.

\subsubsection{DER Operational Scenarios}

For both the uniform and clustered DERs, we consider the three operational scenarios discussed in Section \ref{sec:setup_der}:\\[+2pt]
    \underline{\emph{Base case:}} The DERs are out-of-operation and do not provide any power. This case is represented by the original grid in Fig \ref{fig:rop_result} a) and d).\\[+2pt]
    \underline{\emph{Home Microgrid:}} The DERs power homes at the node where it is located, 
    contributing to a reduction in the net load as described by \eqref{eq:load_reduction}. The resulting net load for the uniform and clustered DER cases is shown in Fig. \ref{fig:rop_result} c) and f), respectively. \\[+2pt]
    \underline{\emph{Community Microgrid:}} The DERs 
    provide power to customers both at the node where it is located and to neighboring nodes, and are modeled as generators. The DERs are indicated by light blue dots in  Fig. \ref{fig:rop_result} b) and e).\\[+2pt]

\begin{table}[]
    \centering
    \small
    $$
    \begin{array}{ll}
    \toprule \text {\textbf{Parameter} } & \text {\textbf{Node/line numbers} } \\
    \hline 
    & \\[-7pt]
    \text {Uniform DERs} & 50,42,56,2,27,5,26,45,22,43,\\& 25,4,
    52,28,12,8,37,12,14,17,\\& 25,6,22,36,28,42,28,12,41\\
    \hline 
    & \\[-7pt]
    \text {Clustered DERs } & 40,41, 42,43,44,45,46,47,48,\\& 49,50, 51,52,53,54,55 \\
    \hline 
    & \\[-7pt]
    \text {Damaged Lines } & 2,10,24,43,23,47,28,19,7,35,\\& 40,33,6,14,42,17,13,50\\
    \bottomrule
    \end{array}
    $$
    \caption{Parameters for modified IEEE 123 test case.}
    \label{tab:parameters}
\end{table}

\subsubsection{Implementation}

The optimization problems are defined in the \emph{PowerModelsRestoration.jl} \cite{rhodes2020powermodelsrestorationjl} software package and implemented in the Julia programming language \cite{julia}.  The ROP is solved using Gurobi v9.1 \cite{gurobi} and the RIP is solved 
using Ipopt \cite{wachter2006implementation}.  
The experiments were conducted on a desktop with an Intel 10-core 5GHz processor and 16 GB of RAM.

\section{Numerical Results} \label{results}
We start by discussing the restoration ordering solutions that we obtain from the ROP,
before analysing how sensitive those solutions are to inaccuracy in either the power flow formulation or the assumptions regarding DER operation.

\subsection{Restoration Ordering Problem Solution}
\begin{figure*}[t]
    \centering
    \includegraphics[width=1.0\textwidth]{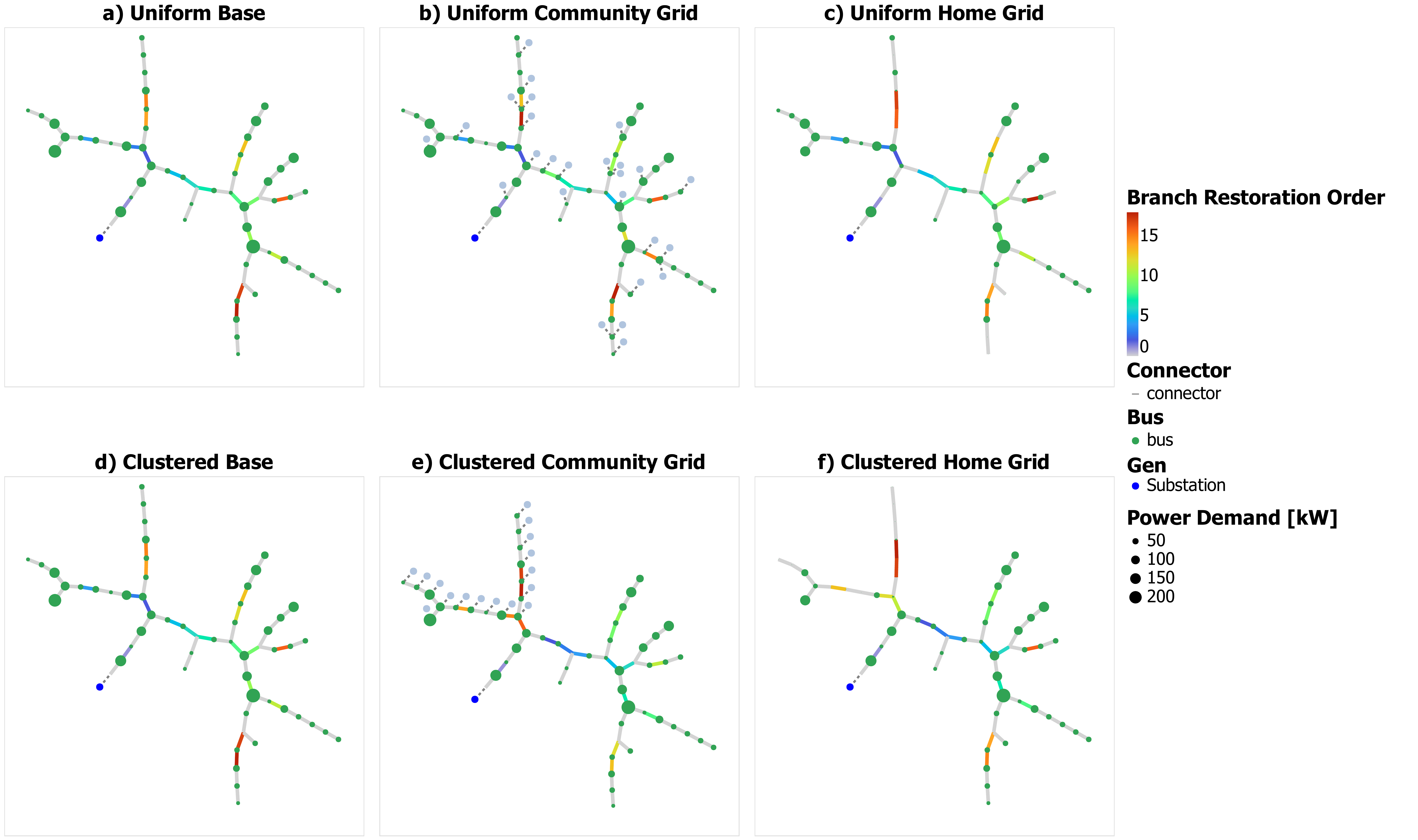}
     \caption{\small The order of repaired lines for different DER operational scenarios and different geographical distributions.  
     The substation is indicated by the dark blue circle, and the DER locations are indicated by light blue circles in the community microgrid case. The size of the nodes (in green) are scaled to indicate the power demand, with larger circles representing larger loads.
     Gray lines is used to represent undamaged power lines. Damaged lines are drawn with a color scale that represents the order of repairs, with blue as the first repairs, then green, yelow and finally red.
     The top three networks a), b) and c) represents the cases with uniformly distributed DERs, while the bottom three networks d), e) and f) are the three cases with clustered DERs. From left to right, the networks reflect the different DER operating scenarios, with the base case (left), the community microgrid (middle) and the home microgrid (right). 
     }
    \vspace{-10pt}
     \label{fig:rop_result}
\end{figure*}

The ROP is first solved for each of the six case study variations shown in Fig. \ref{fig:rop_result}. For each of the solutions, we record (1) the total energy not served in MWh across all time periods, (2) the ordering of the repairs, and (3) the average time of grid reconnection and energy not served for customers with and without DERs. 

\begin{figure}%[h]
    \centering
    \includegraphics[width=0.5\textwidth]{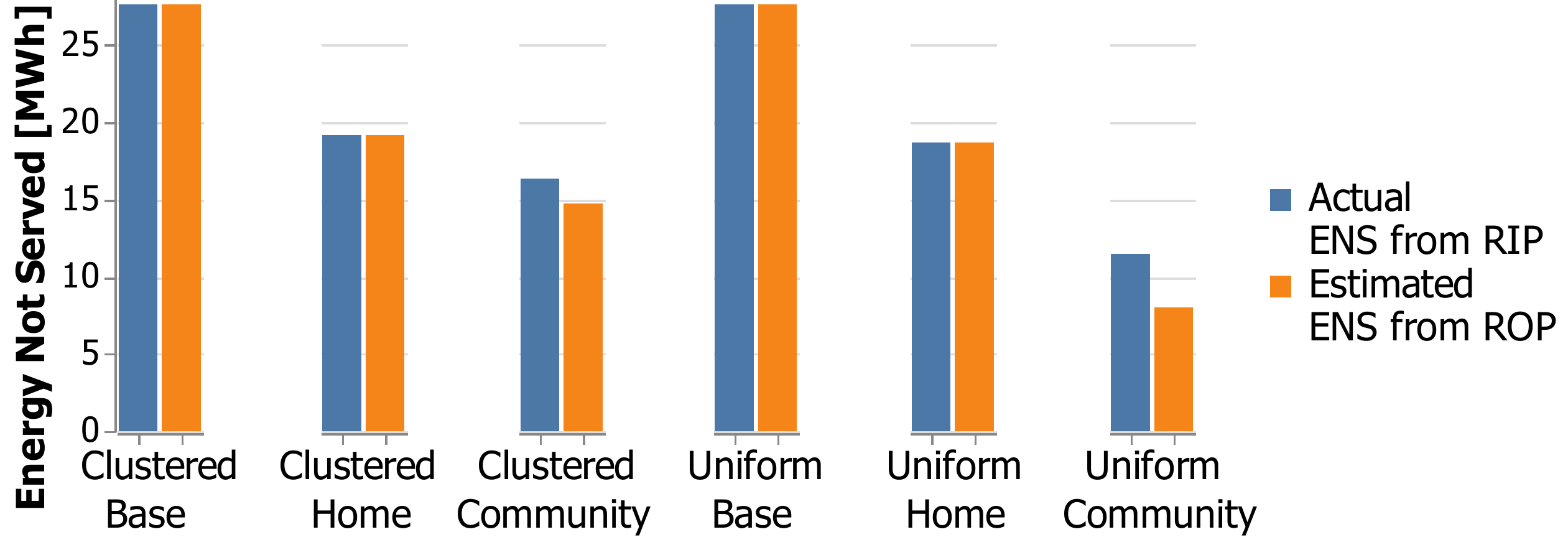}
     \caption{\small Energy Not Served (ENS) for the the DC power flow-based ROP solutions (blue bars) and AC power flow-based RIP solutions (orange bars).}
     \vspace{-10pt}
     \label{fig:dc_accuracy}
\end{figure}

\subsubsection{Total energy not served}

The ROP maximizes the amount of energy served to customers during the restoration process. However, this result can also be used to estimate the total \emph{energy \textbf{not} served} (ENS), which is calculated as
\begin{equation}
    ENS = \sum_{t \in \mathcal{T}}  \sum_{\demandid \in {\demand}} (1 - x_{\demandid,t})   \boldsymbol{D}_{\demandid} \boldsymbol{\Delta}_t.
\end{equation}
For simplicity of comparison, we use $ENS$ as our performance metric.  

The ENS results are shown as blue bars in Fig. \ref{fig:dc_accuracy}. We observe that the base case has the highest $ENS$ 27.7 MWh (more than 40\% of the total demand), followed by the uniform and clustered home grid cases with an $ENS$ of 18.7 MWh (28.2\%) and 19.2 MWh (28.9\%), respectively. The community microgrid cases have the lowest $ENS$ with 11.5 MWh (17.3\%) for the uniform case and 16.4  MWh (24.7\%) for the clustered case.

The difference in ENS can be explained by the access to DER power. 
In the base case, the loads only receive power from the substation, meaning that the energy at a node will remain unserved until reconnected to the substation. 
In the home grid cases, the loads on nodes with DERs are able to receive either full or partial power even before they are reconnected to the grid, which reduces the $ENS$. 
It is interesting to observe that both the uniform and clustered cases have relatively comparable $ENS$ in this case, even though the uniform case has a larger number of DERs. 
This is partially due to the fact that some nodes are not able to make full use of their DER power because they are not able to share excess power with their neighbors. 
This changes in the community grid cases, where the DERs are able exchange power with nearby nodes, leading to a further reduction in $ENS$. 
However, in the clustered DER case, a significant portion of the DER power remained unused because the power can only be exchanged with nearby nodes. 
In the uniform DER case, this limitation is however less severe, as the DERs are spread across a larger area (and there is a larger number of DERs). 
This demonstrates the benefit of having DERs located uniformly across the grid in a disaster scenario. 

\subsubsection{Order of repairs}

We next investigate how the repairs are prioritized in the different cases. The restoration ordering is
shown in Figure \ref{fig:rop_result} where coloring allows us to visually observe the ordering of the repairs. The non-damaged lines are shown in light gray, and the ordering of the repairs for the damaged lines is indicated by the colors. Blue lines are repaired first, followed by green, yellow and red lines.

In the base case, shown in Fig \ref{fig:rop_result} a) and c), repairs are conducted from the substation towards the largest loads in the network. The first four repairs reconnect the large loads on the left side of the grid, and then the next few repairs reconnects the central feeder on the right side. Finally, the more peripheral parts of the feeder are restored, ordered by the largest load.

In the cases with uniformly distributed DERs, the initial repair order remains mostly the same as in the base case. In the uniform home grid case, shown in Fig. \ref{fig:rop_result} c), the initial repairs occur from the substation towards the largest loads. However, the reduction in power demand due to the DERs changes the priority of some of the lines.
In the uniform community microgrid case, shown in Fig \ref{fig:rop_result} b), the first several repairs are also the same as the base grid. However, the next few repairs expand the community microgrids to serve load using local resources before reconnecting to the substation. 

In the cases with clustered DERs, the solutions differ more significantly from the base case solutions. 
For the clustered home grid case in Fig \ref{fig:rop_result} f), the DERs have significantly reduced the power demand in the left half of the feeder, so the ROP prioritizes repairs to the right side of the grid first. 
All repairs are chosen to reconnect loads to the substation. 
For the clustered community grid case in Fig \ref{fig:rop_result} e), the ROP again first prioritizes repairs to reconnect the loads without any DERs. However, the ROP problem next chooses to repair lines within the clustered DER region to create a local microgrid before reconnecting to the substation. 

These results highlight three aspects of DERs during restoration. 
First, then the objective is to maximize the amount of energy served to customers, nodes powered by DERs are given a lower priority of reconnection to the substation. 
Second, if DERs operate in community microgrid mode, the restoration process is sped up by expanding local microgrids before reconnecting to the main grid. 
Third, the optimal restoration order is more sensitive to the presence of DERs if they are concentrated in a limited region of the network.

\subsubsection{Average Reconnection Time and ENS}

\begin{figure}%[h]
    \centering
    \includegraphics[width=0.5\textwidth]{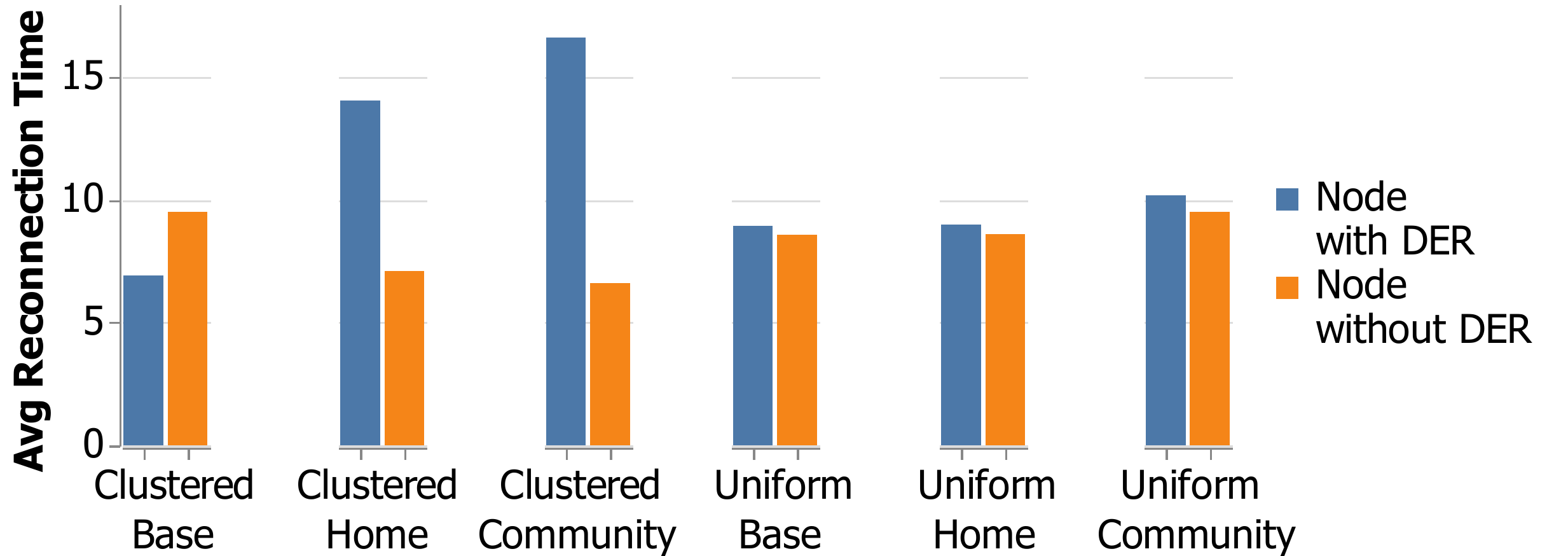}
     \caption{\small The average reconnection time for nodes with and without DERs. The plot shows the reconnection time for the base case, home microgrid and community microgrid with clustered DERs (left) and  uniform DERs (right).}
     \label{fig:ttr}
\end{figure}

We next investigate the average reconnection time and total ENS for loads connected at nodes with and without DERs. We first consider the average reconnection time, as described by \eqref{eq:reconnection}. 
Figure \ref{fig:ttr} shows the average time to reconnect a customer load to the substation for nodes with DERs (blue) and without DERs (orange). The base case results are different in the uniform and clustered cases, because the locations of the DERs are different. 

In the uniform DER case, the reconnection time is not very sensitive to the DER operational scenario, but is slightly higher for nodes with DER across all scenarios. In the clustered DER case, the choice of DER operational scenario has a significant impact on reconnection time for the different groups of customers. In the clustered base case, the nodes with DERs are connected three hours \emph{earlier} on average when compared to nodes without DERs. In the clustered home and community microgrid cases, the nodes with DERs are reconnected on average 7 and 10 hours \emph{later}. 

\begin{figure}%[t]
    \centering
    \includegraphics[width=0.5\textwidth]{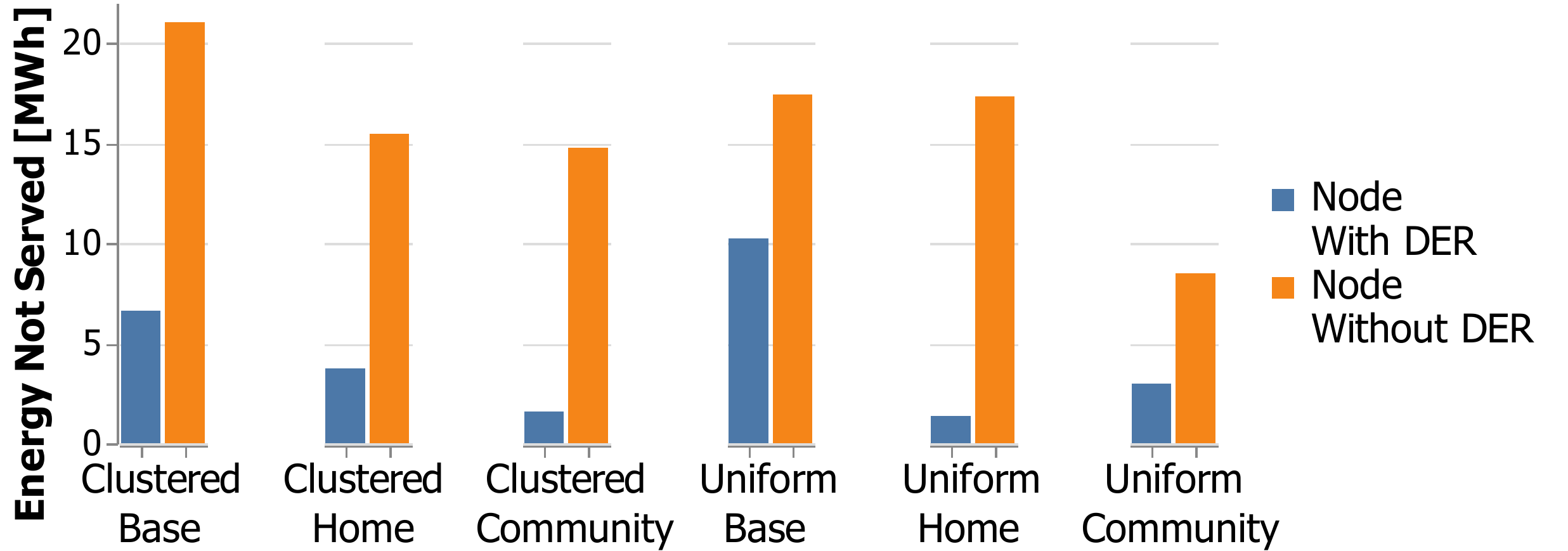}
     \caption{\small The \emph{ENS} for nodes with and without DERs. The plot shows the \emph{ENS} for the base case, home microgrid and community microgrid with clustered DERs (left) and  uniform DERs (right). }     
     \vspace{-10pt}
     \label{fig:ens_by_der}
\end{figure}

Next, we compare the total ENS for nodes with and without DERs, as shown in Fig. \ref{fig:ens_by_der}. We observe that the ENS is higher at nodes without DERs. It is worth noting that this holds  for all cases, including in the base case where the DERs are not providing power. 
Further, we note that the ENS decreases for all nodes when the DERs are operated in home or community microgrid mode. This shows that an increased reconnection time for nodes with DERs in the clustered DER microgrid cases does not translate into a higher ENS.


\subsection{Results of Restoration Implementation}

We next investigate the sensitivity of our solutions to the two main sources of inaccuracy that affect the ROP solution, (1) inaccuracy due to the use of the simplified DC power flow formulation which ignores voltages and reactive power and (2) inaccuracy due to wrong assumptions about DER operational modes. 

\subsubsection{Inaccuracy due to power flow formulation}

We first examine whether the ROP problem provides an accurate estimate for the $ENS$, despite using a simplified DC power flow representation. To assess this, we solve the RIP problem for each restoration order $z$ obtained from the ROP solutions and compare the estimated $ENS$.  The results are shown in Fig. \ref{fig:dc_accuracy} with the solution from the ROP in blue and the actual $ENS$ from the RIP in orange.

For the base scenario and the home microgrid scenarios, the ROP and RIP result is exactly the same ENS. This is because there are no binding power flow or voltage constraints in those cases, the non-damaged line connections (which are accurately represented in the ROP) are the only constraints that matter. In the community microgrid case, the reactive power from the DER resources is not sufficient to support the loads. As a result, the RIP requires additional load shed within the community microgrids, leading to a higher $ENS$.

\subsubsection{Inaccuracy due to DER operational scenario}
\begin{figure}%[h]
    \centering
    \includegraphics[width=0.5\textwidth]{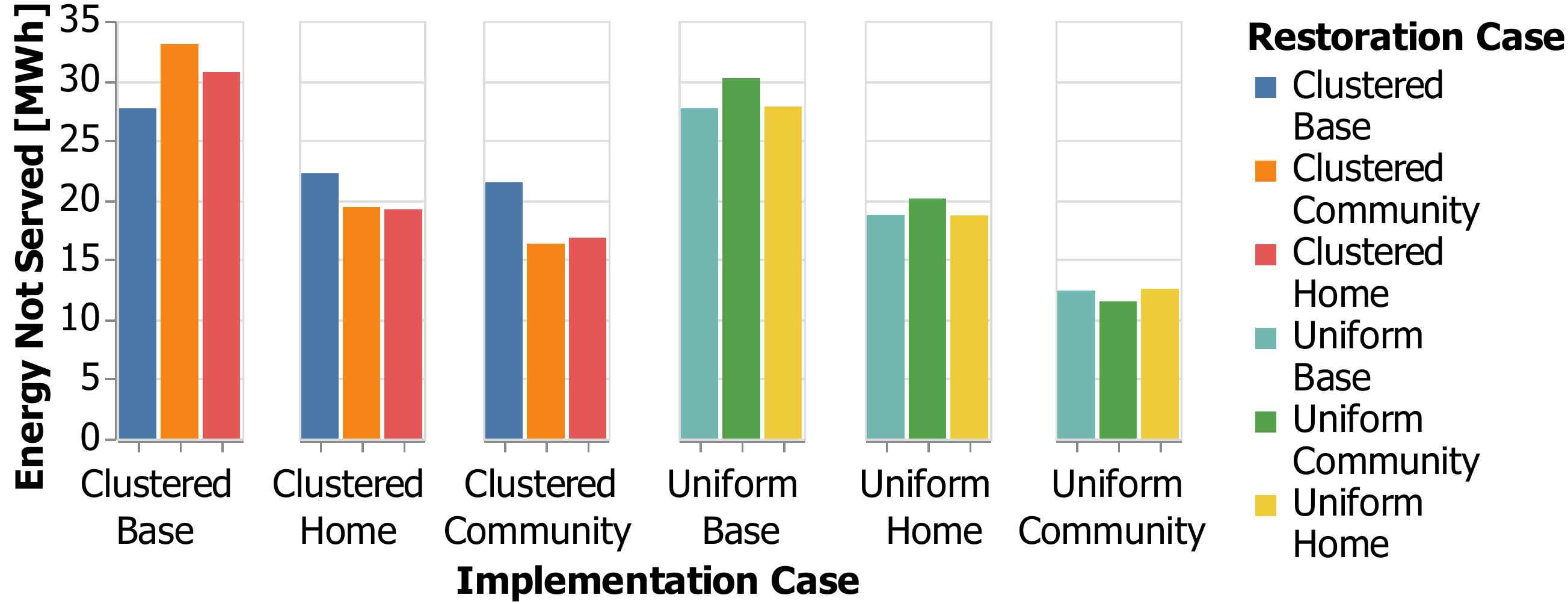}
     \caption{\small The sensitivity of ENS to inaccurate assumptions regarding DER operations. Results for clustered DERs are shown on the left and uniform DERs on the right.}
     \vspace{-10pt}
     \label{fig:dispatch_sens}
\end{figure}

The restoration plan might be implemented under different conditions than the ones that were assumed when the ROP was solved.  
To assess the sensitivity of the solution to the assumptions regarding DER operation, we again start with the different restoration ordering solutions from the ROP. For each solution, we solve three RIP problems assuming that the DERs are either (i) unavailable, (ii) operating as a home microgrid, or (iii) operating as a community microgrid. Given that we have three ROP solutions and each has three corresponding RIP solutions, we get a total of nine cases each for the clustered and uniform DER scenarios. The ENS for each case is shown in Fig. \ref{fig:dispatch_sens} with results for uniform DERs (top) and clustered DERs (bottom). Each group of bars correspond to a different DER operating scenario in the RIP, and the color of the bars in each group represents the ROP problem that was the source of the restoration sequence.    

Our first observation is that the final $ENS$ values are more sensitive to the DER operational assumptions in the RIP than the ROP. This is indicated by the fact that the variation in $ENS$ \emph{between} the groups of bars is higher than the variation \emph{within} each group. 
Our second observation is that the lowest bar in each of the group occurs when the DER operational assumptions are the same in the ROP and RIP problems. This shows the benefit of having accurate knowledge about DERs when preparing and implementing a restoration plan.
Finally, we also notice that the clustered DER case is more sensitive to inaccurate DER assumptions (i.e., has relative large differences in the $ENS$ within each group), as compared to the uniform DER case (where all bars within a cluster are relatively similar in height).

\section{Conclusion} \label{conclusion}

This work studies the optimal restoration sequence for a distribution grid with high penetration of DERs, as well as how the presence of DERs impacts the outage times and energy not served to different groups of customers. 

We discuss a two-step method that first solves a DC ROP problem to identify the optimal restoration plan, and then simulates the implementation of this plan using the AC RIP problem. This process is applied under consideration of three different DER operational scenarios, 0) No DER presence, 1) Community microgrids and 2) Home microgrids. We also consider two different geographical distributions of DERs, 1) Uniform DERs and 2) Clustered DERs. 

We apply our method to a scenario where a windstorm has caused significant damage to a distribution feeder.
We observe that the presence of DERs significantly decreases the \emph{ENS} in the restoration process across all scenarios, for all customers, and that DERs with community microgrid capabilities provide the most benefit.  Clustered DERs significantly change the optimal restoration order, as DER heavy regions have a lower priority to reconnect to the substation.

The current paper leaves several avenues for future work. For example, we would like to include a full three-phase power model, a discrete load shedding and consideration of switching capabilities to isolate the damaged regions, which might make it more difficult for community microgrids to form.

\bibliographystyle{ieeetr}
\bibliography{references}

\end{document}